\documentclass[aps,prd,a4paper,superscriptaddress,nofootinbib,10pt, two column]{revtex4-1}
\usepackage{amsmath,amssymb,graphicx,multirow,dcolumn,bm,latexsym,soul,nicefrac,booktabs}
\usepackage[colorlinks,linkcolor=blue,citecolor=blue,urlcolor=blue]{hyperref}
\newcounter{RomanNumber}

\allowdisplaybreaks
\newcommand{\be}{\begin{equation}}
\newcommand{\ee}{\end{equation}}
\newcommand{\bea}{\begin{eqnarray}}
\newcommand{\eea}{\end{eqnarray}}

\newcommand{\SPA}{School of Physics and Astronomy, Sun Yat-sen University, 2 Daxue Road., Zhuhai 519082, China}

\begin{document}

\title{Constraints on Brans-Dicke gravity from neutron star-black hole merger events using higher harmonics}

\author{Jing Tan}
\affiliation{\SPA}

\author{Baoxiang Wang}
\email{wangbx25@mail2.sysu.edu.cn}
\affiliation{\SPA}

\date{\today}

\begin{abstract}
Brans-Dicke (BD) theory is one of the simplest scalar-tensor theories, with potential applications in dark matter, dark energy, inflation, and primordial nucleosynthesis. The strongest constraint on the BD coupling constant is provided by the Cassini measurement of the Shapiro time delay in the Solar System. Constraints from gravitational wave (GW) events are subject to asymmetric binaries. The third Gravitational-Wave Transient Catalog (GWTC) reports a neutron star-black hole (NSBH) merger event, GW200115, making it possible to constrain BD by GW. With the aid of this source and Bayesian Markov-chain Monte Carlo (MCMC) analyses, we derive a 90\% credible lower bound on the modified parameter of scalar-tensor theories as $\varphi_{-2}>-7.94\times10^{-4}$ by using dominant-mode correction. Specific to BD theory, we have the constraint $\omega_{\rm BD}>4.75$. Asymmetric binary systems usually have a significant mass ratio; in such cases, higher harmonic modes cannot be neglected. Our work considers higher harmonic corrections from scalar-tensor theories and provides a tighter constraint of $\varphi_{-2}>-7.59\times10^{-4}$. Transitioning to the BD theory, the constraint is $\omega_{\rm BD}>5.06$, with a 6.5\% improvement. We also consider a plausible NSBH event, GW190814, which is a highly unequal mass ratio source and exhibits strong evidence for higher-order multipoles. We obtain poorly converged results when using the dominant mode while getting a constraint of $\varphi_{-2}>-6.60\times10^{-4}$ on scalar-tensor theories when including the higher harmonic modes. This suggests that the difference between the dominant mode and higher modes has a significant impact on our analysis. Furthermore, treating this suspected event as an NSBH event, we find $\omega_{\rm BD}>6.12$ when including the higher harmonic modes. Combining GW200115 and GW190814 and including higher modes, the constraint is improved to $\omega_{\rm BD}>110.55$. This is currently the strongest constraint utilizing GWs, contingent upon GW190814 being an NSBH event. Additionally, we take into account a BD-like theory, known as screened modified gravity (SMG), and investigate the coupling constant constraints, both with and without higher-mode corrections, by using data from both GW200115 and GW190814.
\end{abstract}
\maketitle

\section{Introduction}
Recently, Advanced LIGO  \cite{LIGOScientific:2014pky} and Virgo \cite{VIRGO:2014yos} finished three observations $(O1-O3)$, and their results were reported in the corresponding Gravitational-Wave Transient Catalogs (GWTC1-3), which contain 90 gravitational wave events in total \cite{LIGOScientific:2016aoc, LIGOScientific:2018mvr, LIGOScientific:2020ibl, LIGOScientific:2021usb, KAGRA:2021vkt}. These events have been widely used in astrophysics \cite{LIGOScientific:2020kqk}, cosmology \cite{LIGOScientific:2019zcs}, and general relativity (GR) \cite{LIGOScientific:2016lio, LIGOScientific:2018dkp, LIGOScientific:2019fpa, LIGOScientific:2020tif, LIGOScientific:2021sio, Isi:2019aib, Isi:2017equ}--such as testing the no-hair theorem \cite{Isi:2019aib}, the polarization of GW \cite{Isi:2017equ}, and graviton mass \cite{LIGOScientific:2016lio}. Compared to previous tests, like laboratory and Solar System experiments, or binary pulsar and cosmological observations, the GW events are powerful utilities for testing GR in strong or dynamical fields, where we crucially distinguish general relativity in modified gravity. An increasing amount of GW events are to be implemented to constrain modified gravity, as in dynamical Chern-Simons gravity \cite{Yunes:2016jcc, Nair:2019iur, Perkins:2021mhb, Okounkova:2019zjf, Okounkova:2021xjv}, Einstein-dilaton-Gauss-Bonnet gravity \cite{Yunes:2016jcc, Tahura:2019dgr, Nair:2019iur, Perkins:2021mhb, Wang:2023wgv}, and scalar-tensor theories \cite{Yunes:2016jcc, Zhao:2019suc, Niu:2021nic}.

Scalar-tensor theories \cite{Berti:2015itd, Damour:1992we, sotiriou2014gravity}, one of the most natural extensions of GR, include some scalar degrees of freedom in the gravitation sector of the theory from nonminimal coupling. Such scalar fields non-minimally coupled with gravity can be produced through compactification from higher-dimensional theories, such as string theory \cite{polchinski1998string}, Kaluza-Klein-like theories \cite{duff1994kaluza}, or braneworld scenarios \cite{Randall:1999ee, Randall:1999vf}. Scalar-tensor theories provide a robust framework for examining the phenomenological aspects of various potential fundamental theories; they have applications in studying the accelerating expansion of our Universe \cite{Riazuelo:2001mg, Brax:2004qh, Kainulainen:2004vk}, inflation \cite{Clifton:2011jh, Burd:1991ns, Barrow:1990nv}, structure formation \cite{Brax:2005ew}, and primordial nucleosynthesis \cite{Coc:2006rt, Damour:1998ae, Larena:2005tu, Torres:1995je}. The simplest scalar-tensor theory is BD theory \cite{Brans:1961sx}, which was proposed by Jordan, Fierz, Brans, and Dicke, and constructed from Mach's principle. In BD theory, the coupling constant $\omega_{\rm BD}$ is from the scalar field coupling metric field and is assumed to be an invariable constant. The value of the coupling constant $\omega_{\rm BD}$ determines how much BD theory is modified from GR, and there have been numerous efforts to constrain $\omega_{\rm BD}$. The typical constraints come from the secular change measurements in the compact binary's orbital period \cite{eardley1975observable, Will:1977wq, Will:1989sk, Damour:1992we, Damour:1998jk, Alsing:2011er, Antoniadis:2013pzd, Zhang:2019hos, Seymour:2019tir, Freire:2012mg}. Including the measurement on the orbital decay of the pulsar-white dwarf binary PSR J1738+0333, the constraint on the BD coupling constant is obtained as $\omega_{\rm_{BD}}>25000$ \cite{Freire:2012mg}. The most stringent constraint is $\omega_{\rm_{BD}}>40000$ \cite{Bertotti:2003rm, Will:2014kxa}, which comes from the Cassini measurement of the Shapiro time delay in the Solar System \cite{Bertotti:2003rm}. Meanwhile, there are also many other applications for GW in constraints. GWs in the BD waveform are concentrated by neutron star-black hole (NSBH) binaries because waveform calculations require differences in the ``sensitivities'' of the binary systems \cite{Freire:2012mg}. Therefore, the current constraints on BD are based on NSBH binaries, such as constraining BD using space-based GW detectors in future simulations \cite{Berti:2004bd, Yagi:2009zm, Scharre:2001hn, Jiang:2021htl, Gao:2022hsn}. In the case of using observed GW events for constraints, Rui Niu $et~al$. used the GW200115 event to obtain the result $\omega_{\rm BD}>40$ \cite{Niu:2021nic} through Bayesian inference implemented with the open-source software Bilby \cite{Ashton:2018jfp}. Using the breathing scalar mode, Takeda $et~al.$ report the constraint $\omega_{\rm BD}>81$ \cite{Takeda:2023wqn}.

The available sources for constraining BD are restricted to asymmetric binary systems (binaries with different ``sensitivity'' such as NSBH, white dwarf-neutron stars, or white dwarf-black hole binaries). Such binary systems typically have a relatively large mass ratio, such as GW200115. Furthermore, higher-harmonic-mode effects cannot be neglected for such large mass ratio binary systems \cite{London:2017bcn}. However, the sources used in the works above considered only the dominant $(\ell,|m|=2,2)$ mode and possibly introduced errors because higher modes contain a wealth of information. The absence of higher modes implies the loss of this information, leading to biases in parameter estimation and extraction of important physics. Our work extends the waveform to the higher harmonic mode. Specifically, we use the open-source software PyCBC \cite{Biwer:2018osg} to implement Bayesian inference \cite{Thrane:2018qnx, Smith:2021bqc, Lyu:2022gdr} in the parametrized post-Einsteinian framework (ppE) \cite{Yunes:2009ke, Yunes:2010qb, Mirshekari:2011yq, Tahura:2018zuq} for analyzing GW events. We adopt the waveform template IMRPhenomXHM as our waveform model in this process. From analyzing GW data of GW200115, we obtain the constraint on the coupling constant of scalar-tensor theories with $\varphi_{-2}>-7.94\times10^{-4}$, which is compatible with the constraints provided by the LIGO-Virgo-KAGRA Collaboration (LVK) \cite{LIGOScientific:2021qlt}. With this result, we derive constraints on BD theory, yielding $\omega_{\rm BD}>4.75$. When considering the contribution including higher harmonic waveform template corrections, we find $\varphi_{-2}>-7.59\times10^{-4}$ and $\omega_{\rm BD}>5.06$ from GW200115. Meanwhile, we analyze GW190814, which features a significantly unequal mass ratio, and confirm higher multipole radiation at high confidence \cite{LIGOScientific:2020zkf}. The results show that we cannot obtain well-converged posteriors on scalar-tensor theories when using corrections to the dominant mode alone. In contrast, when corrections include higher modes, we find the constraint $\varphi_{-2}>-6.60\times10^{-4}$ and the result $\omega_{\rm BD}>6.12$ regarding GW190814 as an NSBH event. With the higher-mode results, when combining GW200115 and GW190814, the constraint results have been significantly improved: $\varphi_{-2}> -4.76\times10^{-5} \text{ and }\omega_{\rm BD}>110.55$.  Furthermore, we consider the SMG, which is a theory similar to BD. By using GW200115 and GW190814, we obtain constraints of $\frac{\varphi_{\rm_{VEV}}}{M_{\rm_{Pl}}}<3.3\times10^{-2}$ and $\frac{\varphi_{\rm_{VEV}}}{M_{\rm_{Pl}}}<3.1\times10^{-2}$ when including higher modes. The former shows a 3.0\% improvement compared to the dominant-mode case. Analogously, the combined constraint on SMG is enhanced: $ \frac{\varphi_{\rm_{VEV}}}{M_{\rm_{Pl}}}<8.32\times10^{-3}$.

The rest of this paper is organized as follows: Sec. \ref{sec:model} contains two subsections, with the first one briefly reviewing BD theory and the derivation of GW, and the second focusing on the ppE framework of waveforms. Sec. \ref{sec:Bayes} discusses the data-processing methods, i.e., Bayesian inference. In Sec. \ref{sec:result}, we present the main results, and finally, in Sec. \ref{sec:sumout}, we provide a summary and outlook. We use the convention $G = c = 1$ throughout the paper.

\section{Wavefrom}\label{sec:model}
In this section, we introduce the waveform in BD theory and map it into the ppE formalism.
\subsection{Waveform in Brans-Dicke theory}
BD gravity \cite{Brans:1961sx} is a reduced theory of scalar-tensor theory \cite{Harrison:1972enf,Berti:2015itd, Chiba:1997ms}, which includes some scalar degrees of freedom in the gravitation sector from nonminimal coupling [from Ricci-scalar multiplying a scalar field(s) function]. In Jordan's frame, the action of the massive BD theory can be written as\cite{Brans:1961sx, Alsing:2011er}

\begin{equation}
\begin{split}
  S=&\frac{1}{16\pi}\int d^4x\sqrt{-g}\left[\phi R-\frac{\omega_{\rm BD}}{\phi}g^{\mu\nu}
    \left(\partial_\mu\phi\right)\left(\partial_\nu\phi\right)
    -U(\phi)\right] \\
    &+S_M[\Psi,g_{\mu\nu}]\,,\label{BD_bd_action}
    \end{split}
\end{equation}
where $\phi$ is the scalar field and $\Psi$ denotes the matter fields. Our work focuses on the $U(\phi)=0$ case. From the quadratic action, one can obtain the scalar field equation (see Ref.\cite{Liu:2020moh}) and tensor field equation
\begin{equation}
\square_\eta\theta_{\mu\nu} = -16\pi\tau_{\mu\nu}
\end{equation}
Here 
\begin{equation}
\theta_{\mu\nu}=h_{\mu\nu}-\frac12 h\eta_{\mu\nu}-\frac{\varphi}{\phi_0}\eta_{\mu\nu},
\end{equation}
where $g_{\mu\nu}=\eta_{\mu\nu}+h_{\mu\nu}$ is the decomposition of the metric into Minkowski metric plus a small perturbation and $\phi=\phi_0+\varphi$ stands for the scalar field decomposing into a constant background value plus a small one. Here, $\square_\eta=\eta^{\mu\nu}\partial_\mu\partial_\nu$ is known as the d'Alembertian operator. Using Green's function and multipole expansion, the quadrupole formula of the tensor wave can be obtained as:
\begin{equation} 
\theta^{ij} = \frac{2G(1-\xi)}{D_L}\frac{d^2}{dt^2}\sum_A m_A x_A^ix_A^j,
 \label{BD_feq_qua}
\end{equation}
where $\xi=1/(2\omega_{\rm BD}+4)$, and $D_L$ is the distance from the field point to the center of mass. Here, $m_A,~x_A^i, \text{ and }  x_A^j$ are the mass and coordinates of body $A$. Six polarization modes can be derived from the geodesic deviation equation in the approximation of the long wavelength and low speed of test particles. These include plus, cross, breathing, longitudinal, $x,$ and $y$ polarization. By combining this propagation effect with the generation effect [i.e, Eq.\eqref{BD_feq_qua}], aided by the Riemann tensor of linear order metric perturbation $h_{\mu\nu}$, the expressions for the plus polarization $h_+$ and the cross polarization $h_\times$ can be obtained. In the frequency domain, one has \cite{Liu:2020moh}:
\begin{eqnarray}
\tilde{h}_+(f) &=&-\delta\frac{(G\mathcal{M})^{\frac56}}{D_L} \frac{1+\cos^2\iota}{2}\big(\frac{5\pi}{24}\big)^\frac12 (\pi f)^{-\frac76}\Delta \cdot e^{i\Psi_+}, \nonumber\\
\tilde{h}_\times(f)& =&-\delta\frac{(G\mathcal{M})^{\frac56}}{D_L} \cos\iota\big(\frac{5\pi}{24}\big)^\frac12 (\pi f)^{-\frac76}\Delta \cdot e^{i\Psi_\times},
\label{BD_hcha}
\end{eqnarray}
where $\delta=(1-\xi)^{5/3}[1+\alpha(1-2s_1)(1-2s_2)]^{2/3}$, with $s_1, s_2$ the sensitivities of the binary. Here, $\mathcal{M}$ is the chirp mass determined by the mass of binary components, $m_1$ and $m_2$, and it is expressed as $\mathcal{M}={(m_1m_2)}^{3/5}/{M}^{1/5} ~(\text{ with } M=m_1+m_2)$. Note that $\iota$ is the angle between the detector’s line and the binary orbital angular momentum. The phase $\Psi_\times=\Psi_++\frac{\pi}{2}$ is defined in  \cite{Liu:2020moh} and
\begin{equation}
\begin{split}
\Delta =& 1+\frac56\xi-\frac13\alpha(1-2s_1)(1-2s_2) \\
&-\frac{5}{64}\xi \times
\left[\frac{16}{15}\Gamma^2  + \frac43\frac{\mathcal{S}^2}{(GM\pi f)^\frac23}\right],
    \end{split}
\end{equation}
where $\alpha=1/(2\omega_{\rm BD}+3)$ and $\Gamma=(m_1(1-2s_2)+m_2(1-2s_1))/M$, while $\mathcal{S}=s_1-s_2$. These formulas are the precursors of the ppE waveform.

\subsection{Parameter post-Einsteinian framework}
Our waveform models were constructed using the ppE framework, a generic parametric approach, proposed by Yunes and Pretorius to capture the effects of modified GR  \cite{Yunes:2009ke, Yunes:2010qb, Mirshekari:2011yq, Tahura:2018zuq}. Early ppE method focused on quasicircular inspirals of comparable mass, nonspinning compact binaries only. Then it was extended to more general situations \cite{Sampson:2013lpa, Huwyler:2014gaa, Loutrel:2014vja, Sampson:2013jpa, Sampson:2014qqa}, and widely applied to observed GW data to constrain the non-GR effects \cite{Tahura:2018zuq, Lyu:2022gdr, Wang:2021jfc, Wang:2023wgv, Nair:2019iur}. Below, we will review the ppE framework and implement such an analysis to BD gravity.

The inspiral waveform of compact binaries in the ppE formalism is composed of GR terms and modified parts \cite{Yunes:2009ke}:
\begin{equation}\label{ppe_h_com}
\tilde{h}(f)=\tilde{h}_{\rm GR}(1+\alpha\, u^a)e^{i \beta u^b}\,.
\end{equation}
Note that $f$ is the frequency, and $(\alpha u^a, \beta u^b)$ are the amplitude and phase corrections from modified gravity. Here $u={(\pi \mathcal{M} f)}^{1/3}$ is the reduced frequency of the inspiral that is proportional to the relative velocity of the binary, and $\mathcal{M}$ is the chirp mass. The ppE formalism in Eq.\eqref{ppe_h_com} reduces to GR when the ppE parameters $(\alpha, \beta)$ vanish.

One can map Eq.\eqref{ppe_h_com} to the basis of spin-weighted spherical harmonics  $ Y_{-s}^{\ell m}(\iota, \varphi) $ to obtain the harmonic decomposition of ppE waveform \cite{Mezzasoma:2022pjb}:
\begin{equation}\label{ppe_higher_harmonic}
	\tilde{h}_{\ell m}(f) = \tilde{h}_{\ell m}^{\rm {GR}}(f) (1+ \alpha_{\ell m} u^{a_{\ell m}})\exp\big[{i\beta_{\ell m} u^{b_{\ell m}}}\big] ,
\end{equation}
where $\ell=2,3,4$ are the positive-frequency independent harmonics. We focus on the phase correction because of the negligibility of the amplitude correction. Hence the ppE waveform reduces to 
\begin{equation}
	\tilde{h}_{\ell m}(f) = \tilde{h}_{\ell m}^{\rm {GR}}(f)\exp\big[i\beta_{\ell m} u^{b_{\ell m}}\big].
	\label{wave_ppe_pha}
	\end{equation} 
In scalar-tensor theories, the dominant-mode correction is
\begin{equation}
\beta_{2, 2}=\frac{3}{128}\varphi_{-2}.
\label{wave_phi_2}
\end{equation}
Here $\varphi_{-2}=-5(\alpha_{\rm A}-\alpha_{\rm B})^2/168$, where $\alpha_{\rm A} \text{ and } \alpha_{\rm B}$ are scalar charges of the components of the binary.
From Eq.\eqref{BD_hcha}, the dominant-mode correction of these ppE parameters in BD theory  can be written as \cite{Freire:2012mg,Tahura:2018zuq, Liu:2020moh}:
\begin{eqnarray}
\beta_{2,2} &=& -\frac{5}{1792}\eta^{2/5}\mathcal{S}^2 \xi \label{ppe_beta}, \\
b &=& 2k-5,
\end{eqnarray}
with $k=-1$, where $\eta=m_1m_2/M^2$ is the symmetric mass ratio, and $\mathcal{S}=s_1-s_2$ with $(s_1, s_2)$ is the sensitivity of the binary components, which is roughly equal to its compactness (0.5 for black holes and 0.2 for neutron stars). The phase correction with higher harmonics in \eqref{wave_ppe_pha} can be derived as \cite{Mezzasoma:2022pjb}:
\begin{equation}\label{ppe_beta_b}
	\beta_{\ell m} = \beta_m , \qquad b_{\ell m} = b ,
\end{equation}
with 
\begin{equation}
	\beta_{m} = \left(\frac{2}{m}\right)^{2(k-4)/3}\beta_{2,2} ,
\end{equation}
where $\beta_{2,2}$ is from Eq.\eqref{ppe_beta}.

\section{Bayesian Inference}\label{sec:Bayes}
\begin{table}[thb]
\centerline{$\begin{array}{c|l}\hline\hline
\mathcal{M}    & \text{$\mathcal{M}=(m_1m_2)^{3/5}/M^{1/5}$ is the chirp mass.}   \\
q        & \text {$q=m_2/m_1(<1)$is the mass ratio of binary.}\\
a_1, a_2        & \text {Dimensionless spin magnitudes.}\\
\theta_1, \theta_2        & \text{Polar angles of the spin angular momentum of binary.} \\
\phi_1, \phi_2           & \text{Azimuthal angles of the spin angular momentum.} \\
\alpha, \delta 		    &  \text{Sky location of the binary.}      \\
\psi      & \text{The GW polarization angle.}\\
\iota     & \text{The orbital plane inclination angle of the binary.}  \\
\phi_{\rm ref}       & \text{The reference phase at the reference frequency.}  \\
t_c              & \text{The coalescence time.}   \\
D_L              & \text{The luminosity distance.}  \\
\omega_{\rm_{BD}}              & \text{The coupling constant of BD.}  \\
\hline\hline
\end{array}$}
\caption{\protect\footnotesize
Summary of physical parameters and their meaning. The chirp mass $\mathcal{M}$ is in the detector frame, and the GW polarization angle $\psi$ is in earth-centered coordinates. While the inclination angle $\iota$ is from orbital angular momentum to the observer's line of sight.}\label{tableI}
\end{table}

\begin{table*}[t]
\caption{Summary of constraints on coupling constants at a 90\% credible level. }
\label{result_summary}
\begin{ruledtabular}
\begin{tabular}{ccccccc}
\multirow{2}*{} & \multicolumn{2}{c}{GW200115} & \multicolumn{2}{c}{GW190814} & \multicolumn{1}{c}{Combination} \\
\cline{2-3}\cline{4-5}\cline{6-6}
& Dominant mode(DM) & Higher mode (HM) & Dominant mode(DM) & Higher mode (HM) & Higher mode (HM) \\
\colrule
  $\varphi_{-2}$ & -7.94e-4 & -7.59e-4  & $\backslash$ & -6.60e-4 & -4.76e-5& \\
  $\omega_{\rm_{BD}}$ & 4.75 & 5.06  & $\backslash$ & 6.12 & 110.55 \\
  $\frac{\varphi_{\rm_{VEV}}}{M_{\rm_{Pl}}}$ & 3.4e-2 & 3.3e-2  & $\backslash$ & 3.1e-2 &  8.32e-3\\
\end{tabular}
\end{ruledtabular}
\end{table*}

\begin{figure}[h!]
\begin{tabular}{c}
\includegraphics[scale=0.565]{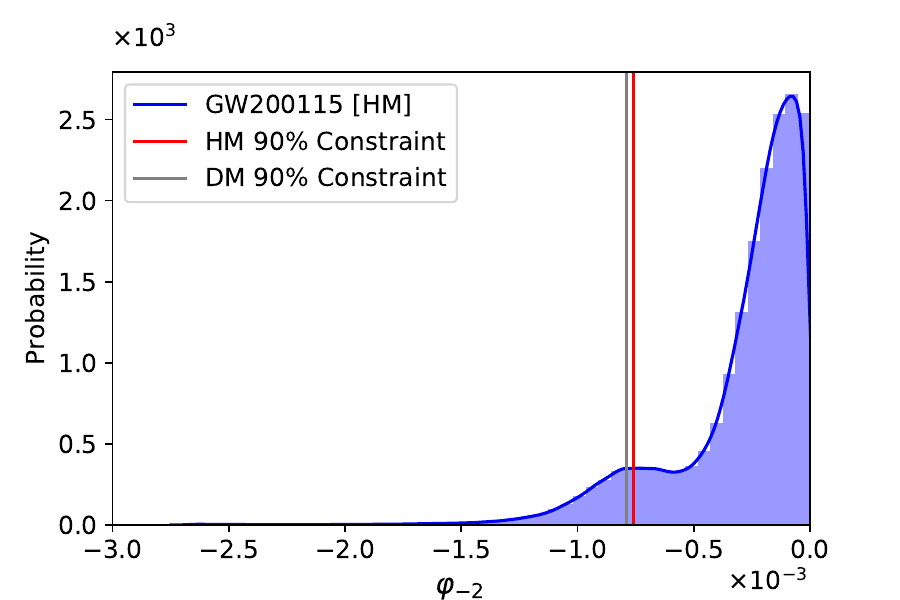}
\\
\includegraphics[scale=0.565]{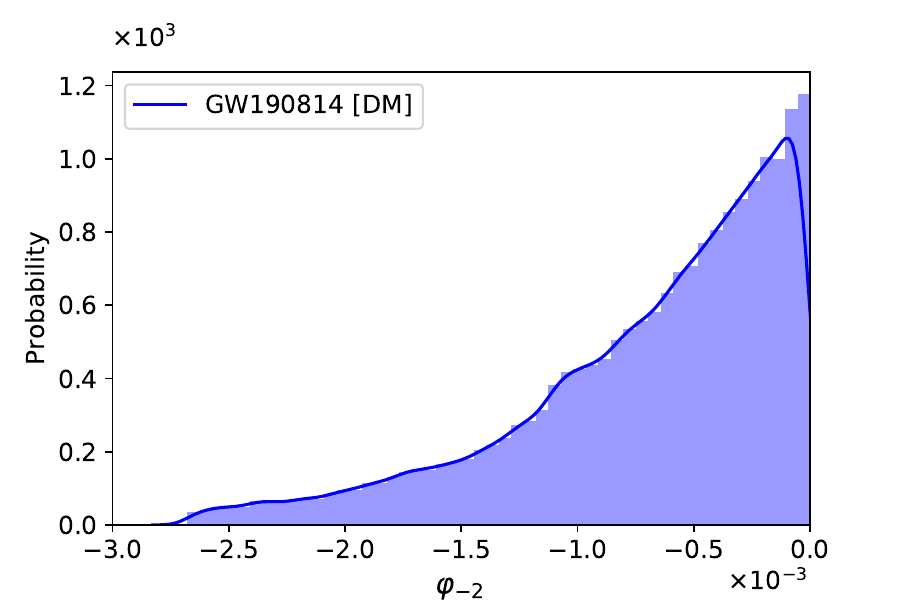}
\\
\includegraphics[scale=0.565]{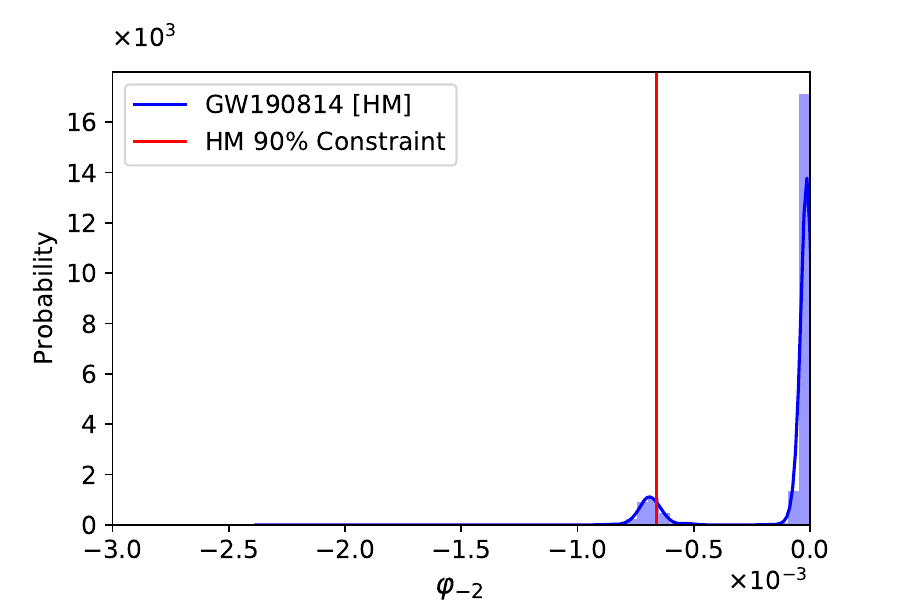}
\end{tabular}
 \caption{Probability density of Bayesian posterior distributions on $\varphi_{-2}$. The gray line corresponds to the constraint at a 90\% credible level by using the dominant-mode (DM) correction, while the red line stands for the constraint at the 90\% confidence level by using the correction including the higher mode (HM). The top panel shows the constraints on $\varphi_{-2}$ from the GW200115 event; the middle and bottom panels are from GW190814, and they show the results corresponding to dominant and higher modes. The absence of a gray line in the middle panel is due to the fact that there is not good convergence in the result when using the dominant mode. We use a logarithmic scale due to the posterior distribution of $\omega_{\rm BD}$ spanning several orders of magnitude.}
\label{fig_phi_2}
\end{figure}

\begin{figure}[h!]
\begin{tabular}{c}
\includegraphics[scale=0.6]{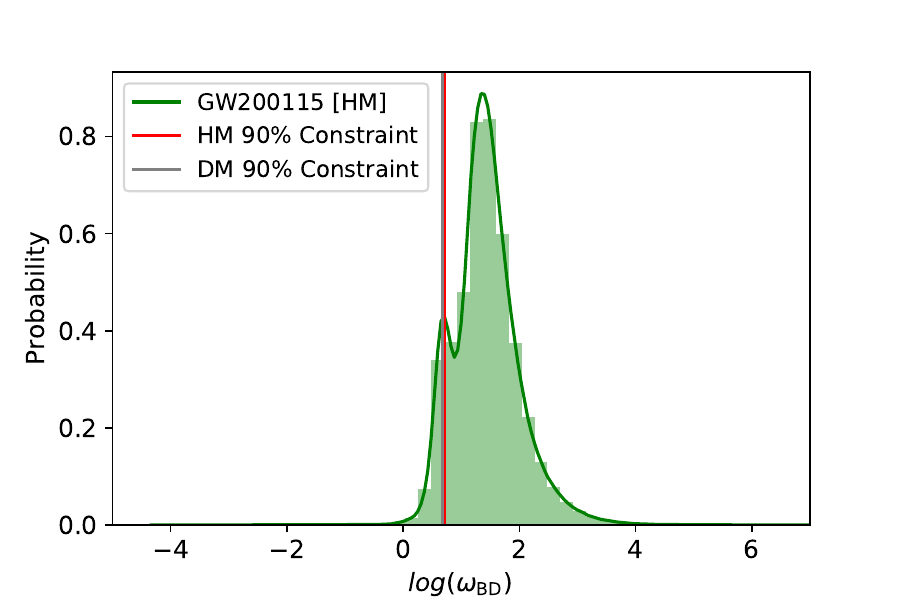}\hspace{0cm}
\\
\includegraphics[scale=0.6]{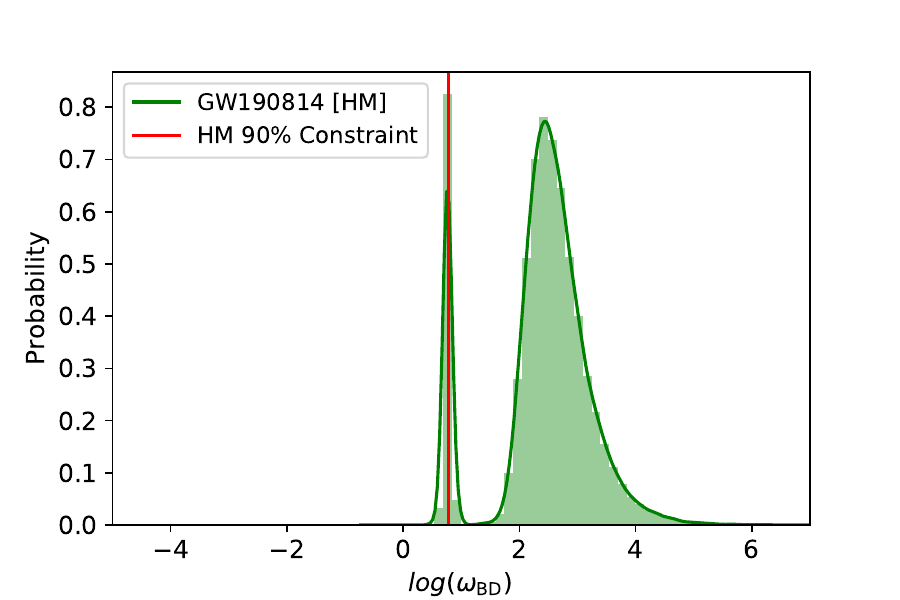}\hspace{0cm}
\end{tabular}
 \caption{Probability density of Bayesian posterior distributions on $\omega_{\rm_{BD}}$. The top and bottom panels describe the constraints on $\omega_{\rm_{BD}}$ from GW200115 and GW190814 events, respectively.  The gray line corresponds to the constraint on BD at 90\% credible level by using the DM correction, while the red line represents the constraint on BD with 90\% probability by using the correction including HM. The reason for the absence of the gray line in the bottom panel is also the poor convergence of the dominant mode.}
\label{fig_bd}
\end{figure}

The ppE waveforms can provide prior information about GWs. Using this prior information, one can extract the desired information from the observed waveforms by Bayesian inference \cite{Thrane:2018qnx, Smith:2021bqc, Lyu:2022gdr}. Suppose one possesses the probabilities of events $A$ and $B$ occurring, as well as the probability of event $B$ occurring given that event $A$ is true; then the probability of event $A$ occurring given that $B$ is true can be directly determined by using Bayes' theorem \cite{stuart1994kendall}:
\begin{equation}
p(A|B)=\frac{p(A)p(B|A)}{p(B)},
\end{equation}
where $p(A) \text{ and }p(B)$ are the probabilities of observing $A$ and $B$, respectively. The conditional probability $p(B|A)$ is interpreted as the likelihood, and $p(A|B)$ is the posterior probability. Bayesian inference inherits the idea of Bayes' theorem. In other words, under a given hypothesis $\mathcal{H}$, suppose one possesses the probability distribution of the parameter $\boldsymbol{\vartheta}$, the probability distribution of the data stream $d$ given some parameter $\boldsymbol{\vartheta}$ [i.e., $p(d|\boldsymbol{\vartheta})$], and the probability of data $d$ [which can be given by taking the sum of $p(d|\boldsymbol{\vartheta}_i)$ in discrete uniform distribution]; then one can obtain the probability distribution of parameter $\boldsymbol{\vartheta}$ from Bayesian inference:
\begin{equation}
\label{bayes_infre}
p(\mathcal{\boldsymbol{\vartheta}} | d, \mathcal{H} ) 
=  \frac{p(\boldsymbol{\vartheta} | \mathcal{H}) \; p(d | \boldsymbol{\vartheta}, \mathcal{H}) }{p(d | \mathcal{H})} 
= \frac{p(\boldsymbol{\vartheta} | \mathcal{H}) \;p(d | \boldsymbol{\vartheta}, \mathcal{H}) }{\int d\boldsymbol{\vartheta}\; p(d | \boldsymbol{\vartheta}, \mathcal{H})\;p(\boldsymbol{\vartheta} | \mathcal{H})},
\end{equation}
where the distribution $p(\boldsymbol{\vartheta} | \mathcal{H})$ is called the prior on $\boldsymbol{\vartheta}$. Note that $ p(d | \boldsymbol{\vartheta}, \mathcal{H})$ is named the likelihood function, and it comes from the calculation of the ppE method. Bayesian evidence $p(d|\mathcal{H})$ is an indispensable part of Bayesian inference, and it can be calculated by summing the likelihood [e.i., $p(d|\mathcal{H})=p(d|\boldsymbol{\vartheta}_1,\mathcal{H})+p(d|\boldsymbol{\vartheta}_2,\mathcal{H})+p(d|\boldsymbol{\vartheta}_3,\mathcal{H})\cdots$] in the case of a discrete uniform distribution. The logarithm of the likelihood function with stationary Gaussian noise can be written as
\begin{equation}
  \log p(d|\boldsymbol{\vartheta}, \mathcal{H}) =  \log\bar \alpha -\frac{1}{2} \sum_{k}\left< d_k - h_k(\boldsymbol{\vartheta}) | d_k - h_k(\boldsymbol{\vartheta}) \right>,
\end{equation}
Where log$\tilde{\alpha}$ is the normalization factor, the index $k$ describes the different detectors, and $d_k$ and $h_k$ are the data and waveform templates. Here, ``$<|>$'' denotes the inner product, which is defined as
\begin{equation}
\label{bayes_inner}
  \left<a(t)|b(t)\right> = 2\int \frac{\tilde{a}^*(f)\tilde{b}(f) + \tilde{a}(f)\tilde{b}^*(f)}{S_n(f)} df,
\end{equation}
where $S_n(f)$ is power spectral density (PSD). Note that $a(t)$ and $b(t)$ are signals in the time domain while $(\tilde{a}(f), \tilde{b}(f), \tilde{a}^*(f),  \tilde{b}^*(f))$ are signals in the frequency domain.

Using Bayesian inference, we analyze 16 waveform parameters, including the BD coupling constant $\omega_{\rm_{BD}}$:\\
$\boldsymbol{\vartheta}=(\mathcal{M}, q, a_1, a_2, \theta_1, \theta_2, \phi_1, \phi_2, \alpha, \delta, \psi, \iota, \phi_{ref}, t_c, D_{L}, \omega_{\rm_{BD}})$
The meanings of various parameters are summarized in Table \ref{tableI}.

\section{Results}\label{sec:result}
The Brans-Dicke correction to the phase of the general relativity waveform depends on the binary system's sensitivity difference. Therefore, the same sensitivity of symmetric binaries gives rise to a vanishing correction of BD to GR \cite{Freire:2012mg}. The currently confirmed GW event caused by the asymmetric binary is only GW200115 from GWTC-3. Thus, we consider the event GW200115 to investigate the influence of higher-mode corrections by calculating Bayesian posterior distributions on parameters $\boldsymbol{\vartheta}$, especially $\omega_{\rm BD}$. In addition, we also utilize the plausible NSBH event, GW190814, to implement our study on scalar-tensor theories. Then we can think of GW190814 as an NSBH coalescence event to constrain BD. GW190814 shows powerful evidence of higher multipoles and is a suitable source for studying higher-mode effects. Thus, the correction with higher modes in Bayesian inference makes our calculation more credible. Our Bayesian inference is implemented through the open-source software PyCBC \cite{Biwer:2018osg} by MCMC sampling and the emcee\_pt sampler \cite{Foreman-Mackey:2012any} set by 200 walkers. To avoid excess noise in low frequencies, we apply a cutoff in frequency at $f_{\rm {low}}=20$ for all GW events. Meanwhile, we analyze 64 s of data for GW200115 and 32 s of data for GW190814. As for priors, we choose a uniform distribution for $1/(\omega_{\rm{BD}}+2)$ in the range [0, 0.5] and for $\varphi_{-2}$ in the range [-0.0027, 0]. In terms of the spin setting case, we use isotropic spin distributions with magnitude $a_1, a_2\le0.99$ in all our analyses.

We use the GWs from the inspiral stage and neglect the tidal effects of NSBH \cite{LIGOScientific:2021qlt} in our work. Hence, we choose the IMRPhenomXHM model from the LALSimulation package \cite{collaboration2018ligo} for both the GR and BD waveform models. These waveforms are phenomenological and operate in the frequency domain, accounting for spin precession and higher-order multipole radiation modes.

We present the constraints on $\varphi_{-2}$ in scalar-tensor theories as shown in Fig.\ref{fig_phi_2}. From GW200115, with the dominant-mode correction, the 90\% confidence lower bound of the posterior distribution for $\varphi_{-2}$ is $\varphi_{-2}>-7.94\times10^{-4}$. This result is compatible with the LVK constraint on $\varphi_{-2}$, and the value changes to $\varphi_{-2}>-7.59\times10^{-4}$ when the higher modes are included in the correction, representing a 4.4\% enhancement. On the other hand, using GW190814, we obtain an unsatisfactory convergence when the correction is the dominant mode; see the middle panel of  Fig.\ref{fig_phi_2}. By contrast, when using the correction containing the higher modes, we obtain the constraint $\varphi_{-2}>-6.60\times10^{-4}$ with 90\% probability. We consider the combination of these two events as well. For the misconvergence above, the combined constraint only involves a higher-mode correction case. When using higher modes the combined constraint exhibited notable enhancement: $\varphi_{-2}> -4.76\times10^{-5}$. 

With the constraints of the scalar-tensor theories, we further obtained constraints on $\omega_{\rm BD}$ in BD as shown in Fig.\ref{fig_bd}. When applying the dominant-mode correction, the lower bound of the posterior distribution for the BD from GW200115 indicates $\omega_{\rm BD}>4.75$ with a 90\% credible level. In contrast, the inclusion of higher modes in the correction leads to a 90\% bound of $\omega_{\rm BD}>5.06$ from GW200115, resulting in an improvement of 6.5\%. Analogously, suppose GW190814 is an NSBH coalescence event; then we find a 90\% credible level constraint on BD of $\omega_{\rm BD}>6.12$ when the correction includes higher modes. Similar to the above, the vanishing of dominant modes is also due to poor convergence. Actually, we do not consider another possible NSBH event, GW190426\_152155, again due to convergence issues. When considering the combination of these two events with the higher modes, the constraint has also significantly improved: $\omega_{\rm BD}>110.55$. This stands as the current best constraint on BD using GWs, provided that GW190814 is indeed an NSBH event.

Note that our analysis does not take into account the breathing mode. However, Takeda $et~al.$ report a tighter constraint on BD, with $\omega_{\rm BD}>81$, by including the breathing mode in their analysis \cite{Takeda:2023wqn}. When their analysis excludes the breathing mode, the result is consistent with the work of  Niu $et~al.$, where $\omega_{\rm BD}>40$, also without the breathing mode \cite{Niu:2021nic}. Nevertheless, when converting this result into a constraint on $\varphi_{-2}$, it differs from the constraints provided here as well as by LVK \cite{LIGOScientific:2021qlt}. The reason may be that they use a different noise power spectral density (PSD) than LVK and the one we use here. Both Takeda and Niu use the available event-specific (e.g. GW200115) PSD in LVK posterior sample releases, while here and in the work of LVK, the PSD from strain data is estimated using the Welch method, the latter most closely resembling a real event (see the section about PSD in \cite{PyCBCLink}). Another possible influencing factor is the use of different upper cutoff frequency $f_{\rm max}=f_{\rm ISCO}=(6^{3/2}\pi m)^{-1}$. The total mass $m$ used by Takeda and Niu is from LVK posterior sample releases, while, LVK's and our $f_{\rm ISCO}$ is not a fixed number but varies among different MCMC realizations. In other words, they use the total mass given in GR to calculate the cutoff frequency in modified gravity. These might be the reasons why results from Takeda and Niu are consistent, but different from the results of LVK and us.

We also implement this higher-order effect in a BD-like theory named SMG theory. SMG is one of the scalar-tensor theories that take into account scalar-field self-interaction characterized by potential $U(\phi)$. Here we have the scalar charges of a neutron star and black hole in Eq.\eqref{wave_phi_2} \cite{Zhang:2017srh, Niu:2021nic}
$$\alpha_{\rm A}= \frac{\varphi_{\rm VEV}}{M_{\rm Pl}\Phi_{\rm A}}, ~~\alpha_{\rm B}=0,$$
where $M_{\rm Pl}=\sqrt{1/8\pi G}$ and $\Phi_{\rm A}=Gm/R$ is the surface gravitational potential. Here, we choose $m=1.4M_\odot\text{ and } R=10{\rm km}$. Based on the same process, we constrain the SMG by using GW200115 and GW190814; we summarize the findings in Table \ref{result_summary}. In the case of GW200115, we find the constraints on the coupling constants as $\frac{\varphi_{\rm_{VEV}}}{M_{\rm_{Pl}}}<3.4\times10^{-2}$ and $\frac{\varphi_{\rm_{VEV}}}{M_{\rm_{Pl}}}<3.3\times10^{-2}$ by using the dominant mode and when including higher modes, respectively.  The corresponding higer-mode result for GW190814 is $\frac{\varphi_{\rm_{VEV}}}{M_{\rm_{Pl}}}<3.1\times10^{-2}$. Similarly, the absence of the dominant-mode result is due to the nonconvergence. Considering the combination, we obtain a tighter constraint as well, $\frac{\varphi_{\rm_{VEV}}}{M_{\rm_{Pl}}}<8.32\times10^{-3}$.

\section{SUMMARY AND OUTLOOK}\label{sec:sumout}
In this work, we reviewed the GW calculation in BD theory and the waveform in the ppE framework. Based on Bayesian inference, we then employed the MCMC to analyze the GW200115 and GW190814. By using these two events, we explored the constraints on BD with the correction, including and excluding the higher modes.

Since constraints on BD require the use of an NSBH binary and the number of confirmed NSBH events is currently very limited, future GW detectors, including ground-based and space-based detectors, as well as next-generation detectors, will involve more sources. With more NSBH GW events expected to be detected in the future, we can provide a more comprehensive constraint. We also anticipate the emergence of sources with larger mass ratios in the future, as higher-order effects will be more pronounced in sources with higher mass ratios, allowing for more accurate waveform computation. Regarding waveform accuracy, we only considered scalar dipole radiation and did not account for scalar quadrupole radiation or corrections from scalar mass. In the future, considering these factors will lead to more reasonable waveforms, enhancing the persuasiveness of our results.

\acknowledgments
J.T. would like to thank Chong-Bin Chen and Hong-Xiang Xu, as well as Niu Rui for useful discussions. This research has made use of data, software, and/or web tools obtained from the Gravitational Wave Open Science Center (https://www.gw-openscience.org), a service of LIGO Laboratory, the LIGO Scientific Collaboration, and the Virgo Collaboration.

\bibliography{TQ-BD}
\end{document}